# Binomial Tree Model for Convertible Bond Pricing within Equity to Credit Risk Framework

K. Milanov

PhD student at Institute of Mathematics and Informatics, Bulgarian Academy of Sciences

and

O. Kounchev

Institute of Mathematics and Informatics, Bulgarian Academy of Science, & IZKS- University of Bonn

In the present paper we fill an essential gap in the Convertible Bonds pricing world by deriving a Binary Tree based model for valuation subject to credit risk. This model belongs to the framework known as Equity to Credit Risk. We show that this model converges in continuous time to the model developed by Ayache, Forsyth and Vetzal [2003]. To this end, both forms of credit risk modeling, the so-called *reduced* (constant intensity of default model for the underlying) and the so-called *synthesis* (variable intensity of default model for the underlying) are considered. We highlight and quantify certain issues that arise, as transition probability analysis and threshold values of model inputs (tree step, underlying stock price, etc.). This study may be considered as an alternative way to develop the price dynamics model of Ayache et al. [2003] for convertible bonds in credit risk environment.

## 1  Introduction

In the present paper we fill an essential gap in the Convertible Bonds pricing world by deriving a Binary Tree based model for valuation subject to credit risk.

The literature that presents the valuation framework of convertible bonds in terms of security contingent on the underlying stock and subject to credit risk modeling begins with a quantitative strategies research note of Goldman Sachs of 1994, [6] and ends with the recent white papers of Bloomberg of 2012, [2], [3]. This framework based on the geometric Brownian motion as a stochastic equity model, has been further developed by Ho and Pfeffer [1996], [9], and by Tsiveriotis and Fernandes [1998], [15]. At the end of the same decade, Davis and Lischka [1999], [5], initiated the modern valuation framework incorporating the credit risk. The straightforward inclusion of the intensity rate in the drift of the equity, zero equity price in the event of default, and the inclusion of a recovery rate makes their models (surveyed by Grimwood and Hodges [2002], [7]) more consistent than each of the preceding.

Lateron, Ayache, Forsyth and Vetzal [2003], [1], elaborated further the assumptions of Davis and Lischka [5], and provided a single-factor framework for valuing risky convertible bonds. Namely, they considered precisely what happens on default with respect to both debt value and equity value, assuming optimal action by the holder of the convertible. Also, they developed a Black-Scholes type partial differential equation that represents *pure price* dynamics[1]

---

[1]"Pure" price does not obey the execution of embedded options as callability, puttability etc., but intrinsically obeys the event of default.

of convertible bond.

Regarding the pricing algorithm, Ayache et al. [2003], [1], developed a numerical technique that is based on finite-difference schemes (FDS). By contrast, Davis and Lischka [1999], [5], suggested to use trinomial trees, although they did not implement the model in their paper.

Let us remind that for many peope and software vendors trained in finance, the binary tree is the most preferable and acceptable numerical technique, mainly due to its transparency and speed. In this context, we fill a gap with respect to implementation of the modern framework within tree methods. The most recent publications based on tree models are the 8-th edition of book of Hull [2011], [11], and the monograph of Spiegeleer and Schoutens, [13].

These presentations miss several important issues on credit default modeling, as derivation of the convertible bond pricing algorithm, and convergence of the numerical algorithms. The above references do not present important details about these modeling aspects, and they have considered only the case of total stock default where the underlying stock drops to zero.

Our main contributions are as follows:

- We develop binary tree pricing algorithm, presenting consistently and in detail all modeling aspects in a more general framework compared with the recent publications.
- In the case of the popular *synthesis* credit risk modeling (see e.g. Muromachi [1999] [12], Takahashi et al. [2001], [14], and Ayache et al. [2003], [1], and related publications), we highlight and quantify a lower threshold bound of stock price below which a given binary tree can not determine in a consistent way the convertible bond value.
- We show that when the step of the binary tree tends to zero, the pure convertible bond value on the proposed binary tree converges to the price model of Ayache et al. [2003].
- We compare our results with some previous publications, in particular, we show that the recently available convertible bond model in [11] is irrelevant.

For simplicity of exposition, we avoid considering various contractual complications such as *call notice periods*, *soft call provisions*, *trigger prices*, etc. Also, we assume that risk-free interest rate term structure is flat and that the underlying stock does not pay dividends. The extension of the model to handle both stocks that pay out dividend and an interest rate as a known function of time can be made in the same way as in the classical binary tree approach.

## 2 The Binary Tree Model Derivation

We develop a model based on a sole state variable $S_t$ describing the price of the underlying stock. We model the default of a company by means of a drop of its equity price, eventually to zero. This framework corresponds to empirical observarions. In particular, Clark and Weinstein [1983], [4], claim that in the considered period the common stock value dropped on default in the average about $30\%$. To this end, we will assume that the return of the underlying stock follows a process that is a combination of diffusion process and a Poisson (jump) process. Namely, in the *risk neutral world* for the underlying stock which does not pay dividend we adopt that its price follows the stochastic process[2]

$$dS_t = (r + \lambda\eta)S_t dt + \sigma S_t dW_t - \eta S_t dq_t \tag{1}$$

---

[2]Equation (1) with $\eta = 0$ is the main contribution of Davis and Lischka [1999]. .

where $dS_t$, $dW_t$ and $dq_t$ are the increments for infinitesimal time period of the stock price, Wiener process and homogeneous Poisson process with intensity $\lambda$, respectively[3]. In addition, we assume that there is no correlation between the Wiener process and the Poisson process. Also, $r$ stands for risk free interest rate, and $\eta$ stands for percentage of the stock fall immediately after default. The latter is valid due to the following statement.

**Proposition 1** *Across the moment of exactly one arrival of the Poission process the value of the process (1) drops with exactly $\eta$ percent.*

Let the arrival time for the Poisson process be $\tau$ and let us put $\tau^+ = \tau + \varepsilon$ where $\varepsilon$ satisfies $0 \leq \varepsilon \leq \varepsilon_1$, where $\varepsilon_1 > 0$ is such that no other arrival has happened in the interval $[\tau, \tau + \varepsilon_1]$. Then we know that $dq_t = \delta(t - \tau)dt$ in the interval $[\tau, \tau + \varepsilon_1]$, where $\delta(t)$ is the Dirac delta function. Hence, from extended Itô's lemma it follows

$$\ln(S_{\tau^+}) - \ln(S_t) = (r + \lambda\eta - \frac{\sigma^2}{2})S_t dt + \sigma S(t)dW_t + \ln(1-\eta)dq.$$

This implies the equality

$$\ln(S_{\tau^+}) - \ln(S_t) = \ln(S^c_{\tau^+}) - \ln(S_t) + \ln(1-\eta),$$

where $S_{\tau^+}$ is the value of the process in just one arrival ($dq = 1$), and $S^c_{\tau^+}$ is the value of the process in absence of arrival ($dq = 0$). Since,

$$\lg \frac{S_{\tau^+}}{S^c_{\tau^+}} = \lg(1-\eta)$$

we arrive at the relation

$$S_{\tau^+} = S^c_{\tau^+}(1-\eta).$$

For example, the sudden $70\%$ fall in the asset price through the default is modeled by putting $\eta = 0.7$.

## 2.1 Random Walk Model of Defaultable Stock

In order to model binomial random walk with possibility of default let us consider equation (1) in time discretization with step $\delta t$. In this discretization we obtain

$$\frac{\delta S_t}{S_t} = (r + \lambda\eta)\delta t + \sigma \delta W_t - \eta \delta q_t, \tag{2}$$

and for the stock returns we have the approximations

---

[3]This means that the increment of this kind of Poisson process over a given time interval with length $\delta t$ obeys the Poisson distribution with parameter $\lambda \delta t$.

$$\mathsf{E}\left(\frac{\delta S_t}{S_t}\right) = (r+\lambda\eta)\delta t - \lambda\eta\delta t = r\delta t$$
$$\mathsf{D}\left(\frac{\delta S_t}{S_t}\right) = \sigma^2\delta t + \lambda\eta^2\delta t = (\sigma^2+\lambda\eta^2)\delta t \tag{3}$$

where $\mathsf{E}$ and $\mathsf{D}$ denote the expectation and the variance, respectively.[4] Now, let us focus on the random variable $\frac{S_{t+\delta t}}{S_t}$. Due to the fact that

$$\mathsf{E}\left(\frac{S_{t+\delta t}}{S_t}\right) = \mathsf{E}\left(\frac{S_{t+\delta t}}{S_t}-1+1\right) = \mathsf{E}\left(\frac{\delta S_t}{S_t}\right)+1$$
$$\mathsf{D}\left(\frac{S_{t+\delta t}}{S_t}\right) = \mathsf{D}\left(\frac{S_{t+\delta t}}{S_t}-1+1\right) = \mathsf{D}\left(\frac{\delta S_t}{S_t}\right)$$

and after substitution with equations (4) we obtain

$$\mathsf{E}\left(\frac{S_{t+\delta t}}{S_t}\right) = 1+r\delta t$$
$$\mathsf{D}\left(\frac{S_{t+\delta t}}{S_t}\right) = (\sigma^2+\lambda\eta^2)\delta t. \tag{4}$$

Now, we are ready to model the random walk of a defaultable stock in binomial tree approximating the dynamics given by equation (2), respectively equation (1). First of all, let us note that for the given time step $\delta t$ the Poisson increment $\delta q_t$ may count more than one arrivals if the length of the step is long enough. To this end, we model the event of default in the period $\delta t$ as the event $\{\delta q_t > 0\}$. The probability of this event is equal to $1-\mathsf{P}(\{\delta q_t = 0\})$. Thus, the probability of default, $p_0$, throughout each time step of the binary tree is equal to $1-e^{-\lambda\delta t}$, i.e.

$$p_0 = 1-e^{-\lambda\delta t}. \tag{5}$$

Another major point in the construction of random walks is the post-default behavior of the stock that we adopt. Namely, we assume that once the stock triggers its value in the event of default by means of $\eta$-percent fall, it never moves further. The latter means that we interrupt the stochastic movement after the state of default. Hence, the kernel of binary tree structure [5] should be extended with an imaginary free node which represents the default value of the stock. The scheme of the extended binary kernel is represented in Figure 1. Here we would like to mention that there is no condition on the stock value at the imaginary node. For example, the situation in which $d = (1-\eta)$ is completely possible, althought this is not the case on the plotted scheme in Figure 1.

---

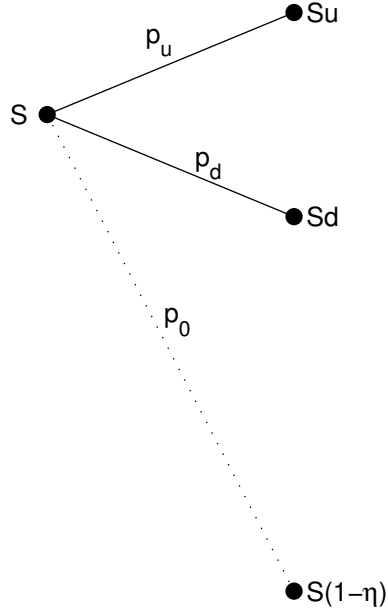

Figure 1: The kernel

Now let us turn to the random variable $\frac{S_{t+\delta t}}{S_t}$. All possible values of $\frac{S_{t+\delta t}}{S_t}$ along the binary tree with time step, $\delta t$ are $u$, $d$, and $(1-\eta)$ with corresponding probabilities $p_u$, $p_d$ and $p_0$, such that $p_u + p_d + p_0 = 1$. Hence,

$$p_d = e^{-\lambda \delta t} - p_u \tag{6}$$

and

$$\mathsf{E}\left(\frac{S_{t+\delta t}}{S_t}\right) = p_u u + \left(e^{-\lambda \delta t} - p_u\right)d + (1-\eta)\left(1 - e^{-\lambda \delta t}\right)$$

$$\mathsf{D}\left(\frac{S_{t+\delta t}}{S_t}\right) = p_u u^2 + \left(e^{-\lambda \delta t} - p_u\right)d^2 + (1-\eta)^2\left(1 - e^{-\lambda \delta t}\right) - \mathsf{E}^2\left(\frac{S_{t+\delta t}}{S_t}\right).$$

Now, using equation (4) we will obtain a system of two equations containing tree parameters, $p_u$, $u$ and $d$.

Regarding $\mathsf{E}\left(\frac{S_{t+\delta t}}{S_t}\right)$, we have the equations:

$$p_u u + \left(e^{-\lambda \delta t} - p_u\right)d + (1-\eta)\left(1 - e^{-\lambda \delta t}\right) = e^{r \delta t}$$

$$p_u(u-d) = e^{r\delta t} - e^{-\lambda\delta t}d - (1-\eta)\left(1-e^{-\lambda\delta t}\right) \tag{7}$$

$$p_u = \frac{e^{r\delta t} - e^{-\lambda\delta t}d - (1-\eta)\left(1-e^{-\lambda\delta t}\right)}{u-d} \tag{8}$$

Regarding $D\left(\frac{S_{t+\delta t}}{S_t}\right)$, we have the equation:

$$(\sigma^2 + \lambda\eta^2)\delta t = p_u(u^2 - d^2) + e^{-\lambda\delta t}d^2 + (1-\eta)^2\left(1-e^{-\lambda\delta t}\right) - e^{2r\delta t}. \tag{9}$$

After replacing with (7) we express all exponentials as series, ignoring terms which contain powers of $\delta t$ bigger than two. Hence,

$$(\sigma^2 + \lambda\eta^2)\delta t =$$
$$= (1 + r\delta t - (1-\eta)\lambda\delta t)(u+d) - (1-\lambda\delta t)ud - (1-\lambda\delta t)d^2$$
$$+ (1-\lambda\delta t)d^2 + (1-\eta)^2\lambda\delta t - (1+2r\delta t)$$

$$(\sigma^2 + \lambda\eta^2)\delta t =$$
$$= (1 + r\delta t - (1-\eta)\lambda\delta t)(u+d) - (1-\lambda\delta t)ud + (1-\eta)^2\lambda\delta t$$
$$- (1+2r\delta t).$$

Due to the fact that we want to build a recombined part of tree, which represents a diffusion, we will look for $u$ and $d$ satisfying $ud = 1$. In addition, let us write $u$ in the form $u = e^{\sqrt{A\delta t}}$. In this way we are able to proceed with the equation of the variance, ignoring all terms containing powers of $\delta t$ bigger than two.

Following the approximation

$$u \approx 1 + \sqrt{A\delta t} + \frac{A\delta t}{2} + \frac{(A\delta t)^3/2}{6} + \frac{(A\delta t)^2}{4!}$$
$$d \approx 1 - \sqrt{A\delta t} + \frac{A\delta t}{2} - \frac{(A\delta t)^3/2}{6} + \frac{(A\delta t)^2}{4!}$$
$$u + d \approx 2 + A\delta t + \frac{(A\delta t)^2}{12}$$

the equation of the variance becomes

$$(\sigma^2 + \lambda\eta^2)\delta t =$$
$$= (1 + r\delta t - (1-\eta)\lambda\delta t))(2 + A\delta t) - 1 + \lambda\delta t + \lambda\delta - 2\eta\lambda\delta t$$
$$+ \eta^2\lambda\delta t - 1 - 2r\delta t$$
$$= 2 + 2r\delta t - 2\lambda\delta t + 2\eta\lambda\delta t + A\delta t - 2 + 2\lambda\delta t - 2\eta\lambda\delta t + \eta^2\lambda\delta t$$
$$- 2r\delta t$$
$$= (A + \lambda\eta^2)\delta t, \tag{10}$$

that is,

$$A = \sigma^2$$

Finally, the parameters of the binomial random model of a defaultable stock, which follows the process (1) are provided in Table 1.

Table 1: Parameters of binomial random walk of a defaultable stock

| Parameter | Definition |
|---|---|
| Multiplier for moving up | $u = e^{\sigma\sqrt{\delta t}}$ |
| Multiplier for moving down | $d = e^{-\sigma\sqrt{\delta t}}$ |
| Probability for moving up | $p_u = \dfrac{e^{r\delta t} - e^{-\lambda\delta t}d - (1-\eta)(1-e^{-\lambda\delta t})}{u-d}$ |
| Probability for moving down | $p_d = -\dfrac{e^{r\delta t} - e^{-\lambda\delta t}u - (1-\eta)(1-e^{-\lambda\delta t})}{u-d}$ |
| Probability of default | $p_0 = 1 - e^{-\lambda\delta t}$ |
| Length of tree step | $\delta t$ |

**Proposition 2** *For each market state represented by parameters $r > 0$, $\sigma > 0$, $\lambda \geq 0$, $0 \leq \eta \leq 1$ and for arbitrary adopted tree step $\delta t > 0$, the corresponding parameter $p_u$ that denotes the probability for an up movements of defaultable stock (1) on binary tree satisfies*
$$p_u > 0.$$

By Table 1, since $u - d > 0$, inequality $p_u > 0$ is equivalent to
$$e^{r\delta t} - (1-\eta) > e^{-\lambda\delta t}(d - (1-\eta)).$$
Since the parameters that determine the stock price dynamics satisfy $r > 0$, $\sigma > 0$, $\lambda \geq 0$, $0 \leq \eta \leq 1$, and the size of the tree step satisfies $\delta t > 0$, we see that $e^{r\delta t} - (1-\eta) > 0$. Hence, the above inequality implies equivalence of inequality $p_u > 0$ to the following one:
$$e^{\lambda\delta t} > \frac{d - (1-\eta)}{e^{r\delta t} - (1-\eta)}.$$
Let us prove the last inequality: Since $d \leq 1$ (see Table 1) we obtain $d < e^{r\delta t}$. This implies
$$d - (1-\eta) < e^{r\delta t} - (1-\eta)$$
which gives, due to $e^{r\delta t} - (1-\eta) > 0$, the following inequality:
$$\frac{d - (1-\eta)}{e^{r\delta t} - (1-\eta)} < 1 \leq e^{\lambda\delta t}.$$

This ends the proof.

**Corollary 1** *For each market state that gives parameters $r > 0$, $\sigma > 0$, $\lambda \geq 0$, $0 \leq \eta \leq 1$, and for arbitrary adopted tree step size $\delta t > 0$, parameter $p_d$ which denotes the probability for a down movement of defaultable stock (1) on binary tree satisfies*
$$p_d < 1.$$

Let us note that the equality $p_u + p_d + p_0 = 1$ implies $p_u + p_d = e^{-\lambda \delta t} \leq 1$, hence, $p_u \leq 1 - p_d$. Now, if we assume $p_d \geq 1$ this will come in contradiction to Proposition 2.

**Proposition 3** *Condition*
$$p_d \geq 0$$
*is necessary and sufficient for the parameters $p_u, p_d$ and $p_0$ to belong to the interval $[0,1]$.*

Since the necessity of the statement is trivial we will proceed now with the sufficiency. Namely, let us assume that $p_d \geq 0$. We will show that parameters $p_u, p_d$ and $p_0$ belong to the interval $[0,1]$. Indeed, from Corollary 1 we obtain that $p_d$ belongs to the interval $[0,1)$.[6] Further, by analogy with Corollary 1 we have $0 \leq p_d \leq 1 - p_u$. Hence, we deduce that $p_u \leq 1$. Combining the latter with Proposition 2 we easily obtain that $p_u$ belongs to the interval $(0,1]$. Finally, from the definition for $p_0$ we deduce that $p_0$ belongs to the interval $[0,1]$.

Finally, we are ready to present the main result by which we can tune the binary tree framework for practical use. Namely, we will show the threshold for the length of the binary tree step for which the derived so far methodology will be consistent with respect to a given market state.

**Theorem 1** *For each market state that gives parameters $r > 0$, $\sigma > 0$, $\lambda \geq 0$, $0 \leq \eta \leq 1$, condition*
$$\lambda \delta t \leq \ln\left(\frac{u - (1-\eta)}{e^{r\delta t} - (1-\eta)}\right)$$
*is necessary and sufficient for the parameters $p_u, p_d$ and $p_0$ to belong to the interval $[0,1]$.*

In order to prove this theorem we will merely show equivalence of this condition with the one of Proposition 3. Indeed, due to $u - d > 0$ and $u - (1-\eta) > 0$, the following equivalences are valid
$$-\frac{e^{r\delta t} - e^{-\lambda \delta t} u - (1-\eta)(1 - e^{-\lambda \delta t})}{u - d} = p_d \geq 0 \Leftrightarrow$$

$$e^{-\lambda \delta t}(u - (1-\eta)) \geq e^{r\delta t} - (1-\eta) \Leftrightarrow$$

---
[6] This denotes the right-open interval not containing $1$.

$$e^{-\lambda \delta t} \geq \frac{e^{r \delta t} - (1-\eta)}{u - (1-\eta)}.$$

By the same reason we have $e^{r \delta t} - (1-\eta) > 0$, and therefore we obtain the following equivalences:

$$-\lambda \delta t \geq \ln\left(\frac{e^{r \delta t} - (1-\eta)}{u - (1-\eta)}\right) \Leftrightarrow \lambda \delta t \leq -\ln\left(\frac{e^{r \delta t} - (1-\eta)}{u - (1-\eta)}\right).$$

This ends the proof.

Below, in Table 2 we have provided the parameters of binomial random walks of a defaultable stock[7]. We have to mention that it is easy to see that in the case of stock that pays continuous dividend yield $D$ we merely need to replace $r$ with $r - D$.

Table 2: Tree Parameters of a defaultable stock that drops to zero

| Parameter | Definition |
|---|---|
| Multiplier for moving up | $u = e^{\sigma \sqrt{\delta t}}$ |
| Multiplier for moving down | $d = e^{-\sigma \sqrt{\delta t}}$ |
| Probability for moving up | $p_u = \dfrac{e^{r \delta t} - e^{-\lambda \delta t} d}{u - d}$ |
| Probability for moving down | $p_d = -\dfrac{e^{r \delta t} - e^{-\lambda \delta t} u}{u - d}$ |
| Probability of default | $p_0 = 1 - e^{-\lambda \delta t}$ |
| Length of tree step | $\delta t$ |

## 2.2 Convertible Bond Pricing Algorithm

The main aim of this section is to present the valuing of convertible bonds within the binomial model for the underlying stock derived so far. Let us proceed with construction of a portfolio at time $t$ that consists of one convertible bond and a short position in a quantity $\Delta$ of the underlying. At time $t$ this portfolio has value

$$\Pi = V - \Delta S,$$

where $V$ stays for the convertible bond value. In time $\delta t$ the convertible bond takes one of three values $V^+$, $V^-$ or $X = \max(RN, k(1-\eta)S)$, depending on whether the underlying stock rises to a value $Su$, falls to a value $Sd$ or drops to the value $(1-\eta)S$. Here $R$ stays for the recovery rate, $N$ stays for convertible bond face value[8], and $k$ stays for the conversion ratio valid at the moment $t+\delta t$ [9].

---
[7]This is the event when the default stock price drops $100\%$, i.e $\eta = 1$.

Here we follow Ayache et al. [2003] to incorporate the option of the holder to convert the bond after announcement of the bankruptcy. Thus, at the moment $t+\delta t$, the possible portfolio values are

$$\Pi_{t+\delta t} = \begin{cases} V^+ - \Delta Su \\ V^- - \Delta Sd \\ X - \Delta S(1-\eta), \end{cases}$$

What we want now is to eliminate the risk of diffusion in portfolio value. That is, we will express the hedge quantity $\Delta$ from the equation

$$V^+ - \Delta Su = V^- - \Delta Sd.$$

In this way, for

$$\Delta = \frac{V^+ - V^-}{S(u-d)}$$

there are only two values of the portfolio: $X - \frac{V^+ - V^-}{(u-d)}(1-\eta)$ and $\frac{V^-u - V^+d}{u-d}$ which arrive respectively with probability, (5) of the event of default in time-step, $\delta t$ and the probability $e^{-\lambda \delta t}$ for non-default in the same time-step. Thereof

$$\mathsf{E}(\Pi_{t+\delta t}) =$$

$$= \frac{V^-u - V^+d}{u-d} e^{-\lambda \delta t} + \left(X - \frac{V^+ - V^-}{(u-d)}(1-\eta)\right)\left(1 - e^{-\lambda \delta t}\right)$$

$$= \frac{e^{-\lambda \delta t}u + (1-\eta)\left(1 - e^{-\lambda \delta t}\right)}{u-d} V^- - \frac{e^{-\lambda \delta t}d + (1-\eta)\left(1 - e^{-\lambda \delta t}\right)}{u-d} V^+$$
$$+ X\left(1 - e^{-\lambda \delta t}\right).$$

Further, the assumption that the risk of default is diversifiable implies:

$$\mathsf{E}(\Pi_{t+\delta t}) = \Pi_t e^{r\delta t},$$

where $r$ is the continuously compounded interest rate. "Diversifiable" means that through the time step $\delta t$ the return of the portfolio is based on risk free interest rate no matter whether default will arrive or not[10]. In this context

$$e^{r\delta t}\left(V - \frac{V^+ - V^-}{u-d}\right) =$$

$$= \frac{e^{-\lambda \delta t}u + (1-\eta)\left(1 - e^{-\lambda \delta t}\right)}{u-d} V^- - \frac{e^{-\lambda \delta t}d + (1-\eta)\left(1 - e^{-\lambda \delta t}\right)}{u-d} V^+ \quad (11)$$
$$+ X\left(1 - e^{-\lambda \delta t}\right).$$

---

[8]This is the most practical case of recovering, refer to the monograph Spiegeleer and Schoutens [2011].
[9]It is possible that the conversion ratio vary in time.
[10]We have adopted risk neutral world.

Proceeding with grouping in the above equation we obtain

$$e^{r\delta t}V = \frac{e^{-\lambda \delta t}u + (1-\eta)(1-e^{-\lambda \delta t}) - e^{r\delta t}}{u-d}V^- - \frac{e^{-\lambda \delta t}d + (1-\eta)(1-e^{-\lambda \delta t}) - e^{r\delta t}}{u-d}V^+$$
$$+ X(1-e^{-\lambda \delta t})$$

and more precisely

$$V = e^{-r\delta t}(p_u V^+ + p_d V^- + p_0 X) \qquad (12)$$

which shows that the way of valuation of the convertibles within the tree (derived so far) is just the same as the well known manner used in the classical binary tree framework.

## 2.3 Including Coupon Cash-Flow

Let us denote with $t_i^c$ the moment at which the bond paid out a coupon amount $c_i$, also let the moment happens inside the tree step $\delta t$. Then, the risk-neutral assumption implies that the amount $c_i e^{r(t+\delta t - t_i^C)}$ must be added to the bond price at the end of the time step in absence of default, i.e.

$$\Pi_{t+\delta t} = \begin{cases} V^+ - \Delta S u + c_i e^{r(t+\delta t - t_i^C)} \\ V^- - \Delta S d + c_i e^{r(t+\delta t - t_i^C)} \\ X - \Delta S(1-\eta). \end{cases}$$

Now, by analogy with the previous section we obtain

$$E(\Pi_{t+\delta t}) = \frac{e^{-\lambda \delta t}u + (1-\eta)(1-e^{-\lambda \delta t})}{u-d}V^- - \frac{e^{-\lambda \delta t}d + (1-\eta)(1-e^{-\lambda \delta t})}{u-d}V^+$$
$$+ X(1-e^{-\lambda \delta t}) + c_i e^{r(t+\delta t - t_i^C) - \lambda \delta t},$$

and the assumption that the risk of default is diversifiable implies

$$e^{r\delta t}V = \frac{e^{-\lambda \delta t}u + (1-\eta)(1-e^{-\lambda \delta t}) - e^{r\delta t}}{u-d}V^- - \frac{e^{-\lambda \delta t}d + (1-\eta)(1-e^{-\lambda \delta t}) - e^{r\delta t}}{u-d}V^+$$
$$+ X(1-e^{-\lambda \delta t}) + c_i e^{r(t+\delta t - t_i^C) - \lambda \delta t}.$$

Hence,

$$V = e^{-r\delta t}(p_u V^+ + p_d V^- + p_0 X) + c_i e^{-r(t_i^C - t) - \lambda \delta t}$$

which shows that the risk neutral present value of the coupon amount at the previous tree level (at the moment $t$) has to be adjusted with the probability for non-default through the time step till the next tree level (at the moment $t + \delta t$).

## 3 Model Convergence

In this section we will prove that the pure convertible bond price of the model derived so far, converges in continuous time to the one modeled by Ayache et al. [2003], [1].

We continue the link established in equation (11), and obtain

$$e^{r\delta t}\left(V - \frac{V^+ - V^-}{u-d}\right) =$$
$$= \frac{V^- u - V^+ d}{u-d}e^{-\lambda\delta t} + \left(X - \frac{V^+ - V^-}{u-d}(1-\eta)\right)\left(1 - e^{-\lambda\delta t}\right), \quad (13)$$

where $V = V(t,S)$ was defined as the convertible bond price by means of the binomial model. Now, let us assume that $V$ is three times differentiable with respect to $S$, where the first and the second derivatives are continuous with respect to $S$, also let us assume that $V$ is continuously differentiable with respect to $t$. Then we can express $V^+ = V(t+\delta t, Su)$ and $V^- = V(t+\delta t, Sd)$ in Taylor series, respectively

$$V^+ = V + \delta t V_t + (u-1)SV_S + (u-1)^2 \frac{S^2}{2}V_{SS}$$
$$V^- = V + \delta t V_t + (d-1)SV_S + (d-1)^2 \frac{S^2}{2}V_{SS}.$$

Hence,

$$V^+ - V^- = (u-d)SV_S + (u-d)(u+d-2)\frac{S^2}{2}V_{SS},$$

that is

$$\frac{V^+ - V^-}{u-d} = SV_S + (u+d-2)\frac{S^2}{2}V_{SS}.$$

Now let us recall that

$$u + d \approx 2 + A\delta t + \frac{(A\delta t)^2}{12},$$

where

$$A = \sigma^2.$$

Hence

$$u + d - 2 = \sigma^2 \delta t + o(\delta t^2),$$

which leads to equation

$$\frac{V^+ - V^-}{u-d} = SV_S + \sigma^2 \delta t \frac{S^2}{2} V_{SS} \quad (14)$$

After all, in a similar way we have

$$V^- u - V^+ d = uV + u\delta t V_t + u(d-1)SV_S + u(d-1)^2 \frac{S^2}{2}V_{SS}$$
$$- dV - d\delta t V_t - d(u-1)SV_S - d(u-1)^2 \frac{S^2}{2}V_{SS}.$$

Now, using condition $ud = 1$ we obtain $u(d-1)^2 = d(u-1)^2$, indeed

$$u(d-1)^2 = u(d^2 - 2d + 1) = udd - 2ud + u = d + u - 2$$
$$d(u-1)^2 = d(u^2 - 2u + 1) = duu - 2du + d = u + d - 2.$$

Thus,

$$V^- u - V^+ d = (u-d)V + (u-d)\delta t V_t - (u-d)SV_S$$

which is equivalent to

$$\frac{V^-u - V^+d}{u-d} = V + \delta t V_t - S V_S \qquad (15)$$

Further, using equations (14) and (15), equation (13) becomes

$$e^{r\delta t} V =$$
$$= (V + \delta t V_t - S V_S) e^{-\lambda \delta t} - \left( S V_S + \sigma^2 \delta t \frac{S^2}{2} V_{SS} \right)(1-\eta)\left(1 - e^{-\lambda \delta t}\right)$$
$$+ \left( S V_S + \sigma^2 \delta t \frac{S^2}{2} V_{SS} \right) e^{r\delta t} + X\left(1 - e^{-\lambda \delta t}\right).$$

In addition, let us expand the exponents in the above equation, and after dropping the higher order terms in $\delta t$ we obtain the equations

$$(1 + r\delta t) V =$$
$$= (V + \delta t V_t - S V_S)(1 - \lambda \delta t) - \left( S V_S + \sigma^2 \delta t \frac{S^2}{2} V_{SS} \right)(1-\eta)\lambda \delta t$$
$$+ \left( S V_S + \sigma^2 \delta t \frac{S^2}{2} V_{SS} \right)(1 + r\delta t) + X \lambda \delta t$$

and

$$V + r\delta t V =$$
$$= V + \delta t V_t - S V_S - \lambda \delta t V + \lambda \delta t S V_S - S V_S (1-\eta)\lambda \delta t.$$
$$+ S V_S + \sigma^2 \delta t \frac{S^2}{2} V_{SS} + S V_S r \delta t + X \lambda \delta t$$

Hence,

$$(r + \lambda) \delta t V = \delta t \left( V_t + (r + \lambda \eta) S V_S + \frac{\sigma^2 S^2}{2} V_{SS} + \lambda X \right)$$

which implies

$$(r + \lambda) V = V_t + (r + \lambda \eta) S V_S + \frac{\sigma^2 S^2}{2} V_{SS} + \lambda \max(RN, k(1-\eta)S).$$

Finally, we obtain that the pure price on the binary tree satisfies the following partial differential equation (PDE)

$$V_t + \frac{\sigma^2 S^2}{2} V_{SS} + (r + \lambda \eta) S V_S - (r + \lambda) V + \lambda \max(RN, k(1-\eta)S) = 0.$$

Since the last equation is the one derived by Ayache et al. [2003], [1], this accomplishes the proof of our statement that our binary tree model converges to the continuous model of [1].

In addition, let us remark that in the case of total stock default modeling, the pure convertible value of the binomial model (derived so far) will satisfy the following PDE

$$V_t + \frac{\sigma^2 S^2}{2} V_{SS} + (r + \lambda) S V_S - (r + \lambda) V + \lambda RN = 0.$$

# 4 Model with Synthesis Form of Credit Risk Modeling

In the modeling until now we have used a constant intensity rate. However, it is more realistic to model intensity rate to increase as the stock price declines. In this way the exogenous nature of default modeling becomes the so-called *synthesis form* due to information incorporated about behavior of the firm's equity price. In the current study we adopt and implement intensity rate model that was considered by Muromachi [1999], [12], Takahashi et al. [2001], [14], Ayache, et al. [2003], [1], namely

$$\lambda(S) = \lambda_0 \left(\frac{S}{S_0}\right)^\alpha, \qquad (16)$$

where $\lambda_0 > 0$ is the estimated intensity rate at $S = S_0$, and $\alpha < 0$. As an applicable guess for $\lambda_0$ one takes often the observable (desirable) credit spread.

However, in practice this model may involve control of the tree step especially for small stock values, where intensity increases. It is easy to see that this will require increasing number of the tree steps, corresponding to Theorem 1. Obviously this is related to extra computation time. Hence, very often in practice we predefine the number of tree steps, respectively we adopt to use a tree with predefined length of step. If this is the case, we will show that the use of a synthesis form of default modeling will impose existence of a lower threshold bound $S^{\text{å}}$, below which, the given binary tree cannot determine in a consistent way the convertible bond value.

**Proposition 4** *For a given tree structure with time step $\delta t > 0$, condition*

$$S \geq S_0 \left[\frac{1}{\lambda_0 \delta t} \ln\left(\frac{u - (1-\eta)}{e^{r\delta t} - (1-\eta)}\right)\right]^{1/\alpha}, \qquad \alpha < 0 < \lambda_0$$

*is necessary and sufficient for the values of the parameters $p_u, p_d$ and $p_0$ to belong to the interval* [0,1].

Applying Theorem 1 we obtain the following inequality:

$$S^\alpha \leq \frac{S_0^\alpha}{\lambda_0 \delta t} \ln\left(\frac{u - (1-\eta)}{e^{r\delta t} - (1-\eta)}\right), \qquad \alpha < 0 < \lambda_0.$$

Hence, from monotonicity of the power function, it follows:

$$S = (S^\alpha)^{1/\alpha} \geq S_0 \left[\frac{1}{\lambda_0 \delta t} \ln\left(\frac{u - (1-\eta)}{e^{r\delta t} - (1-\eta)}\right)\right]^{1/\alpha}.$$

This ends the proof.

Hence, it is easy to see that for a given tree step $\delta t$ the desired lower bound is

$$S^{\text{å}} = S_0 \left[\frac{1}{\lambda_0 \delta t} \ln\left(\frac{u - (1-\eta)}{e^{r\delta t} - (1-\eta)}\right)\right]^{1/\alpha}, \qquad (17)$$

where $\alpha < 0 < \lambda_0$, $0 < r$, $0 < \sigma$ and $0 \leq \eta \leq 1$.

To handle a situation with $S < S^*$ we suggest to make an extension (say, linear, polynomial or spline) of the bond price model taking into account that for the stock $S$ below $S^{\text{å}}$ the behavior of the convertible bond price is very simple (almost linear convergence to the recovery amount).

# 5 Comparison with Previous Work

In 2011 there appeared two binary tree based models for convertible bond valuation subject to default, the one proposed by Spiegeleer and Schoutens [2011], [13], and the one proposed by Hull [2011], [11].

These presentations are lacking important details about credit default modeling, derivation of the convertible bond pricing algorithm as well as the convergence. Also, the above authors, present only the case of total stock default where the underlying stock drops to zero.

The monograph Spiegeleer and Shoutens [2011] contains notions that are slightly misleading. For instance, they define a number $p$ and call it "probability for up movement" on the tree via the expression (6.77, p.110 respectively table 6.2, p.111). However, this does not define a risk-neutral probability $p$, for up movement of the underlying stock, but by our exposition above we have seen that the right expression in the terms of [13] is given by $e^{-\lambda \delta t} p$.

On the other hand, in [11], (pp. 608-610) there is no sufficient details for producing the binomial tree parameters. As we can see from Table 2, the binary tree model in [11] is distinct from our model only with respect to multipliers for up and down movement of defaultable stock. Quoting word for word [11], the adopted process followed by the underlying stock satisfies the following: "It is assumed that the stock follows geometric Brownian motion except that there is a probability $\lambda \Delta t$ that there will be a default in each short period of time $\Delta t$. In the event of a default the stock price falls to zero and there is a recovery on the bond. The variable $\lambda$ is the risk-neutral default intensity..."

The author has obviously tried to model a GBM+Poisson process reflected by the word "except". However the conclusion of the correct formulas for the parameters of the Binomial Tree under such process have been obtained by us in Section 2.1 above. For completeness sake, let us provide a possible way to obtain the parameters of the Binomial Tree available in [11], under the assumptions made there: By analogy with the techniques applied for obtaining equation (9), following the assumptions in [11], for the variance of the variable $\frac{S_{t+\delta t}}{S_t}$ we obtain the following equation

$$\sigma^2 \delta t = p_u (u^2 - d^2) + e^{-\lambda \delta t} d^2 - e^{2r \delta t}.$$

Hence, proceeding in the same manner as in formula (10), we arrive at

$$\sigma^2 \delta t = (A + \lambda) \delta t,$$

that is,

$$A = \sigma^2 - \lambda.$$

The latter determines the parameter for up-move on the tree as

$$u = e^{\sqrt{(\sigma^2 - \lambda) \delta t}}, \tag{18}$$

which is just the same as the one proposed in [11].

Another inconsistency of the model in [11], arises from the factor $\sigma^2 - \lambda$ in the formula (18) for $u$, since in practice $\sigma^2$ and $\lambda$ are very close or even identical. In such situation, the binary tree process will have a very low volatility and will be different from the process adopted in [11]. The effect of this inconsistency is exhibited in the next section.

# 6  An example

We provide an example of a convertible bond which is used in Ayache et al. [2003] [1], that is with terms and conditions that are given in the Table below.

Table 3: **Convertible Bond Terms and Conditions**

| Issue Date | 6-Jan-2009 |
|---|---|
| Maturity Date | 6-Jan-2014 |
| Conversion | 6-Jan-2009 to 6-Jan-2014 into 1 share |
| Call provision | 6-Jan-2011 to 6-Jan-2014 at 110 |
| Put provision | on 6-Jan-2012 at 105 |
| Nominal | 100 |
| Coupon Rate (annual) | 8%, paid out semi-annually |
| Day Count Convention | Act/365 |
| Business Day Convention | Unadjusted |
| Risk-Free Interest Rate | 5% (continuously compounded) |
| Credit Spread | 2% (continuously compounded) |
| Stock Volatility | 20% |
| | |

Here we design a test which shows misestimation of the embedded options by the [11] model, where $\sigma^2$ is close to $\lambda$. Now let us assume $\sigma = 0.25$, $\lambda = 0.062$ and recovery rate $R = 40\%$. In Figure 2 we show the price profile of the bond (from Table 3) that is given by the model in [11], and the profiles of the following two variants of the model proposed in the present study: the case of *constant intensity* rate $\lambda = 0.062$ and the case of *synthesis intensity rate* model data $\lambda_0 = 0.062$, $\alpha = -0.5$ and $S_0 = 50$. We see that the model of [11] shows underestimation of the price of the embedded option. Another important feature of the synthesis model is that it is the only one which shows adequate behavior in the credit risk environment.

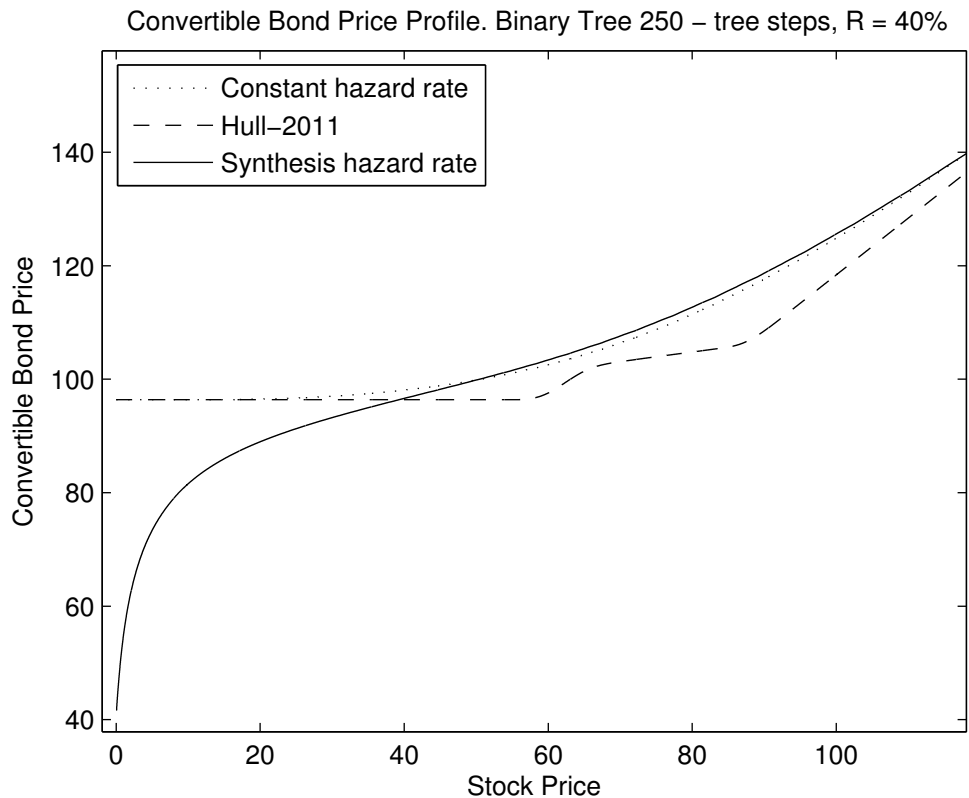

Figure 2: *Mis-estimation of Hull-2011 model in comparison with two typical variants of current work.*

Further experiments will be published in a forthcoming paper.

# References

[1] Ayache, E., P.A. Forsyth, and Kenneth R. Vetzal " Valuation of Convertible Bonds With Credit Risk", *The Journal of Derivatives,* Vol. 11, pp. 9-29, Fall 2003.
[2] Bloomberg, David Frank, "OVCV Model Description. Quantitative Research and Development, Equities Team", January 9, 2012, Version 1.62.
[3] Bloomberg, "Pricing a convertible bond using the Bloomberg jump-diffusion model in OVCV", May 3, 2011, Version: 1.2.
[4] Clark, T.A., and M.I. Weinstein. ``The Behavior of the Common Stock of Bankrupt Firms.'' *Journal of Finance*, 38 (1983), pp. 489--504.
[5] Davis, M., and F.R. Lischka. ``Convertible Bonds with Market Risk and Credit Risk.'' Working paper, Tokyo-Mitsubishi International, 1999.